# Weedy adaptation in *Setaria* spp.: IX. Effects of salinity, temperature, light, and seed dormancy on *Setaria faberi* seed germination

Dekker, J. and J. Gilbert
Weed Biology Laboratory, Agronomy Department, Iowa State University, Ames, Iowa 50011, USA
Email: jackdekker666@gmail.com

**Keywords**: seed dormancy, seed germination, temperature, light, salinity, soil seed banks, salt tolerance, drought tolerance, giant foxtail, weedy foxtails, foxtail species-group

## INTRODUCTION

The ability to grow and reproduce in salty habitats is a function of tolerance to those chemicals at all critical phases of a plant's life history. The ability to withstand salt in the soil during growth and reproduction until it becomes an established vegetative plant may require different mechanisms and plant traits than those needed to overcome dormancy, after-ripen and germinate in the same salty soils. Seeds that are able to establish themselves in soils with high salt content may respond differently depending on the light conditions (soil surface and shallow versus deep burial) and the dormancy states of the seed at the time of salty water imbibition.

The relevance of these questions was most appparent with the discovery and collection of *S. faberi* seed, as well as those from *S. viridis* and *S. pumila* plants, thriving along the seacoasts of Southern Japan, often growing right at the high tide mark of the ocean water of the beach (Dekker, 2000; Dekker, 2014b). These plants had to possess the ability to after-ripen, germinate, emerge and establish themselves, grow and reproduce in the salty soils and salt-laden atmospheres present in these windy habitats.

The world's oceans consists of water and many dissolved solutes, notable various salts (e.g. NaCl). The salt content of seawater is consistent globally at about 3%. Our goal was not to duplicate either the seacoast tidal interface zone, or seawater as an inhibitor, but to conduct a preliminary study about the effects of one important seawater salt to guide future studies on the environmental factors that may interact in the complex processes involved below ground in the life history of weedy *Setaria* seeds in salty environments. In particular the effects of soil temperature, light and the dormancy-germinability states of the seeds during the times of after-ripening in salty soil and the subsequent process of germination.]

Therefore the objectives of this paper are to determine the effect of salt (NaCl) in water imbibed by *Setaria faberi* (giant foxtail) seed during after-ripening and germination, as well as that of temperature and light, in environmentally controlled seed germination conditions. Hopefully these observations may also provide insights on the possible relationship between salt and drought tolerance in the germination process relevant to other more dynamic conditions of climate and more complex sea salt inhibition.

Previous studies in this series have documented the global genotypic population structure of the weedy foxtails (Wang, Wendel, and Dekker, 1995a, b) and some of the agriculturally important phenotypic traits responsible for its extensive biogeographic distribution (Wang and Dekker, 1995; Dekker et al., 1996; Dekker and Hargrove, 2002. Donnelly et al., 2014; Haar and Dekker, 2014; Haar et al., 2014); summarized in Dekker, 2014b.  Weedy adaptation in this species-group also depends on its ability to reproduce in the presence of chemical inhibitors naturally occurring in habitats it has successfully exploited.

## METHODS AND MATERIALS

**Seeds**.  Seed of two populations of *S. faberi* were used to determine the effects of salt (NaCl), temperature and light on germination: non-dormant lot 1816, dormant lot 3781 (Dekker germplasm collection (Dekker, 2014b), Weed Biology Laboratory, Iowa State University).  Non-dormant *S. faberi* population (lot 1816) was derived from a single seed of *S. faberi* originally collected in 1984 from near Woodstock, Illinois (single seed descent scheme described in Dekker et al., 1996).  A single plant was vegetatively separated several times, and the resulting plantlets were transplanted into a field nursery near Ames, Iowa in the growing season of 1991.  The seed of population 1816 were harvested from these panicles and stored in controlled environmental conditions (45-50% relative air humidity (RH), dark, 4°-6° C) in the long-term seed storage facility in the Agronomy Building, Iowa State University, Ames, Iowa, USA until used in these investigations (May-July, 2003).  A dormant *S. faberi* seed population (lot 3781) was also used to determine the effects of germinability-dormancy phenotype and after-ripening on germination in comparisons with non-dormant phenotype lot 1816.  Lot 3781 was collected on October 5, 1999 from a soybean (*Glycine max*) field near Crawfordsville, Iowa.  Immediately after harvest, the seeds were passively air-dried overnight on screens at 19°C.  None of the seed germinated when evaluated at harvest under the conditions described below.  On October 10, 1999 the dried seed was placed in ventilated storage containers and stored at a constant temperature of -20° C. Seeds were removed from these conditions immediately prior to being used in each of the experiments reported here.  The rationale for the choice of populations and phenotypes was based on evaluating treatment effects over a wide range of germinability.  The choice of using fully after-ripened, non-dormant seed of *S. faberi* (lot 1816) was that they possessed the maximum range of germination sensitivity (near-complete germination under ideal conditions) to inhibition by salt, temperature and light.  Dormant seed (e.g. less than 10%; lot 3781) was used to determine responses from a second locally-adapted *S. faberi* population, as well as to determine responses to treatment with after-ripening (in comparison to both fully after-ripened lot 1816 and dormant lot 3781).  Finding a single iso-nuclear genotype with all representative seed germinability phenotypes at the same time is experimentally intractable.  *Setaria faberi* is globally near-homogeneous genetically (allozyme level; Wang et al., 1995b).  Our goal was to describe one locally adapted population of the species in terms of dormant, partially after-ripened and nondormant  phenotypes.

**Germination**.  Germination was evaluated in 30 ml gas-tight vials, with 20 mm outside diameter mouths (Wheaton Science Products, Millville, New Jersey, USA) (as described in Dekker and Hargrove, 2002).  Two disks of Anchor Blue germination blotter paper (Anchor Paper Co., St.

Paul, Minnesota, USA), 32 mm in diameter, were placed in, and completely covered, the bottom of the vials.  Immediately prior to sealing, 1.0 ml of distilled, de-ionized water (or NaCl solutions, as described below) was placed in the vials along with ten dry dark plump healthy-looking mature *S. faberi* seeds.  After placing seeds and water solutions in the vials they were immediately sealed (Wheaton hand crimper, model 22430; Wheaton Science Products, Millville, New Jersey, USA) with neoprene stoppers crimped around the vial neck with an aluminum ring to ensure a gas- and watertight seal.  It was felt that a sealed system was more appropriate because it mimicked the soil seed bank environment in which atmosphere exchange is limited by diffusion processes within the soil matrix, and because it ensured that moisture was at no time either limiting or excessive.  A sealed system also provided us with a known, constant quantity of water that never limited germination.  The amount of water in our assay system was ample for germination and seedling growth up to several leaves.  The seeds were covered at all times with an ample film of water wicked from the blotter paper on which they rested, but were not submerged.  Additional details on the rationale for this type of germination system can be found in Dekker and Hargrove, 2002.  Immediately after preparation, vials were placed in either after-ripening or germination assay conditions.

**After-ripening**.  Seed of population 3781 were after-ripened at 4° C, in the dark (provided by a double layer of aluminum foil around the vials) for 21 days.  Individual vials contained water only (untreated), or water plus the amount of NaCl indicated by their treatment, for the entire after-ripening and subsequent germination assay periods.

**Germination assay**.  The sealed vials with appropriate germination paper, water (and NaCl), and seed were placed in one of three Hoffman (model SG-30, Hoffman Manufacturing, Inc., Albany, Oregon, USA) controlled environment seed germination cabinet for 8 d, after which germination data was collected.  Each of the three chambers was maintained at a single constant temperature of 15°, 25° or 35° C.  In each of these chambers a dark (24h dark:0h light; provided by a double layer of aluminum foil around the vials) or light (15 h light:9 h dark) diurnal light regime was maintained for the duration of the studies.  The fluorescent lighting source in the controlled environment chambers provided light quality and quantity was similar to that reported in Dekker et al., 1996.  Immediately after the eight day germination assay the numbers of seed germinated were determined for each vial.  Germination was evidenced by coleorhiza and/or coleoptile protrusion outside the seed hull (lemma, palea).

**Salt**.  Salt treatments consisted of added amounts of NaCl from 0.01-5.00% (w:v).  Preliminary assays indicated no germination of populations 1816 and 3781 occurred at 2.00-5.00% NaCl.  As such, six levels of salt (0.01, 0.05, 0.10, 0.50, 1.00, 1.50%) were used in comparison with water in the studies reported herein.

**Experimental design and analysis**.  A completely randomized experimental design was used with a factorial arrangement of treatments (parameter, levels), 2 trials (repeats), 12 replications (6 per trial): population and phenotype (3 levels: lot 1816, lot 3781, after-ripened lot 3781); NaCl (7 levels: 0, 0.01, 0.05, 0.10, 0.50, 1.00, 1.50%), temperature (3 levels: 15°, 25°, 35° C) and light (2 levels: light, dark).  Each experiment consisted of 6 replications (vials) per treatment, and the experiment was performed twice (total of 12 replications, 120 seeds, per treatment).  Analysis of variance (ANOVA) was performed on the data and each mean is the pooled mean response of 12 replicates.  Tukey mean separation tests were conducted on the mean final germination percentage for within population-phenotype comparisons only (Table 1, columns) and between population-phenotype comparisons only (Table 1, rows).

## RESULTS

The effects of the interactions among salt, temperature, light and *S. faberi* phenotypes/genotypes were revealed by four types of seed germination comparisons (Table 1): 1] between temperature-light conditions and phenotypes, within a common amount of salt; 2] between different levels of water-salt and phenotypes, at a common temperature and light condition; 3] between light-dark conditions and phenotypes, within a common water-salt amount and temperature condition; 4] between phenotypes within a common water salt amount and temperature-light condition.

**Phenotype/Genotype**

Seed germination of non-dormant 1816 (NON-DORM; 89.8%) was greater than that of dormant 3781 (DORM; 3.9%) and after-ripened 3781 (AFTER-RIPE; 26.3%) when assayed in water only and averaged over all levels of temperature and light (all three means significantly different with Tukey mean separation test).  When averaged over all levels of salt, temperature, and light, germination of NON-DORM (76.8%) was greater than that of DORM (3.2%) and AFTER-RIPE (18.9%) (all three means significantly different with Tukey mean separation test). Overall, germination of dormant population 3781 in all salt, temperature and light conditions was very low (0-11.7%; Table 1), and this may have prevented detection of more subtle effects of salt, temperature and light conditions.  Overall, germination of after-ripened population 3781 ranged from 0-78.3%, while that of non-dormant 1816 ranged from 1.1-98.3%.  These variable responses by all three phenotypes/genotypes were dependent on the particular salt, temperature and light conditions in which they germinated.

**Salt**

Seed germination of dormant and after-ripened population 3781, and non-dormant 1816, were completely inhibited at NaCl amounts of 2.00-4.50% (data not reported).  The effects of 0-1.5% NaCl on germination of these three phenotypes/genotypes were dependent on the specific conditions of temperature and light in which they were incubated and assayed (Table 1).

**Temperature and Light**

The effects of temperature and light on the seed germination of the three phenotypes/genotypes were made evident by comparisons in water-only conditions among the several levels of temperature and light (Table 1).  Seed germination of untreated (water only) NON-DORM was greater than that of untreated DORM in all temperature (15-35° C) or light (light, dark) conditions.  Untreated NON-DORM seed germination was also greater than that of untreated AFTER-RIPE in all but one condition: 35° C in the light.  Water-only seed germination within NON-DORM was similar in all temperature (15-35° C) and light (light, dark) conditions; as was water-only germination within DORM in these same treatments. Comparisons of water-only germination within AFTER-RIPE revealed conditional responses to temperature and light conditions, affects attributable to the process of after-ripening.  After-ripening (4° C for 21 d in dark) *S. faberi* seed population 3781 induced sensitivity to temperature (reduced at low, increased at high) and light (increased in the light).  Water-only seed germination of AFTER-RIPE increased with increasing temperature; in the light, 15-35° C; in the dark from 15° C to 35° C.  Germination of untreated AFTER-RIPE at 25° C was similar in both light conditions, and to that at 35° C in the dark. Water-only seed germination in all three of these temperature-light conditions exceeded that observed at 15° C (light and dark) in which no germination was observed. Germination of untreated AFTER-RIPE was greatest at 35° C in the light, greater than that in any other untreated temperature and light condition for that phenotype. After-ripening of dormant population 3781 increased germination relative to that of DORM within only one common untreated condition: 35° C in the light. Within this single temperature-light condition the increase in germination stimulated by after-ripening was large, 71.1%.

**Salt, Temperature and Light**
Seed of the three phenotypes/genotypes varied in germination response to the presence of salt from 0-1.5% NaCl as a function of the particular temperature and light conditions to which they were exposed. The effects of salt, temperature and light within and among phenotypes/genotypes were revealed by two types of seed germination comparisons (Table 1): among different levels of temperature and light within a common amount of salt; and among different levels of salt at a common level of temperature and light.

**Dormant population 3781**. Seed germination of DORM in only water, and in the presence of NaCl at 1.00-1.50%, was similar in all temperature (15-35° C) and light (light, dark) conditions, when compared within a common level of water-salt. When compared within common temperature and light conditions, germination of DORM was inhibited by NaCl in only one condition, 25° C in the dark at 1.50%. Within all other common temperature and light conditions there was no apparent inhibitory effect of salt on DORM seed germination, possibly because of the overall low germination of this seed accession. The presence of NaCl induced sensitivity to, and germination was dependent on, the light and temperature conditions to which DORM was exposed. There was a tendency for germination to be greater at 25° C in the dark when compared to both higher and lower temperatures, and both light conditions, in common levels of salt (0.01-0.5). Germination of DORM at 25° C in the dark was greater than that at 15° C (light, 0.01-0.5%; dark, 0.1-0.5%) and 35° C (light, 0.05, 0.5%); dark, 0.01-0.05, 0.5%); while germination at 15° C and 35° C was similar when compared at common salt levels. Germination at 25° C in the light was greater than that at 15° C in the light at 0.05% NaCl, but in all other instances germination at 25° C in the light was similar to other temperature-light conditions with a common level of NaCl. Light and dark germination within a common temperature was similar within all common levels of salt.

**After-ripened population 3781**. The process of after-ripening of population 3781 induced changes in germination response to salt temperature and light. The most profound change due to after-ripening was a large increase in germination at 35° C in the light in the presence of 0-0.5% NaCl when compared within common levels of salt. Inhibition of AFTER-RIPE by NaCl was greater than that in DORM in several instances in comparisons with water made at a common level of temperature and light. No AFTER-RIPE seed germinated at 15° C in any amount of NaCl. AFTER-RIPE was inhibited at 25° C (light, 1-1.5%) and 35° C (light 1-1.5%; dark, 1.5%). The only instance of inhibition in DORM (25° C, dark, 1.5% NaCl) was lost in AFTER-RIPE when compared within the same level of temperature and light. Germination of AFTER-RIPE was greatest at 35° C in the light at 0-0.5% NaCl, compared to a tendency for greater germination at 25° C in the dark for DORM, all when compared within the same amount of salt. Germination of AFTER-RIPE was greater at 35° C in the light than in the dark (0-0.5%), and 35° C-light was also greater than that at 25° C (light and dark, 0-0.5%) and 15° C (no germination in any condition) when compared at common amounts of NaCl. Compared to DORM, light-stimulated germination was induced in AFTER-RIPE at 35° C (0-0.5%) and 25° C (0.05%) at common salt amounts, otherwise it was similar at common levels of temperature between 15-25° C in the light and dark.

**Non-dormant population 1816**. Seed germination of untreated, water only, NON-DORM was similar all temperature (15-35° C) and light (light, dark) conditions. NON-DORM germination was largely unaffected by salt, temperature and light with up to 0.1% salt, but was inhibited in all conditions at 1.5% or greater, when compared within common temperature and light conditions. Between 0.01-1.00% NaCl no inhibition occurred at 25-35° C in the light and dark.

The presence of NaCl induced sensitivity to light and temperature conditions in NON-DORM. Inhibition of seed germination at 0.5-1.0% NaCl relative to water-only was observed at 15° C (within common temperature and light conditions); and inhibition was also greater in the light than in the dark (within common NaCl levels) at that lowest temperature. Compared at common salt levels, germination of NON-DORM at 15° C in the light was less than that in the dark (0.5-1.0%), less than that at 25° C in the light (0.05, 0.5-1%) and dark (0.5-1%), and less than that at 35° C in the light (0.05) and dark (0.5-1%). With common amounts of salt exposure, seed germination of NON-DORM increased with increasing temperature from 15-25° C (light, 0.05, 0.5-1%; dark, 1.0%; light and dark, 1.5%), and from 15° C to 35° C (light, 0.05, 0.5-1%; dark, 1%). At 1.5% NaCl germination at 25° C was greater than that at 35° C (light and dark), otherwise responses at these two temperatures were similar.

**Among phenotypes/genotypes**. Germination of NON-DORM was greater than that of either DORM or AFTER-RIPE when compared with a common water/salt (0-1%), temperature and light condition with two exceptions: at 35° C in the light NON-DORM and AFTER-RIPE were similar and greater than that of DORM (0-0.50% NaCl); and at 15° C in the light when NON-DORM and both 3781 phenotypes were similar (1.50% NaCl). At 1.50% salt, NON-DORM had greater germination than the other two 3781 phenotypes at 25° C (light and dark), otherwise germination was similar among populations and phenotypes.

No phenotype was inhibited by the presence of salt from 0.01-0.10%, when compared within common temperature and light conditions. When inhibition of germination by salt did occur, the three phenotypes responded differently. Some of these differences could be a consequence of greater observed inhibition in phenotypes and conditions wherein greater germination occurred, when sensitivity to loss was more apparent.

*S. faberi* seed germination was affected by light in two different situations when compared within common salt concentrations and temperature. The most evident effect was the sensitization of population 3781 germination to light by the after-ripening process. Germination of AFTER-RIPE, unlike DORM, was greater in the light than in the dark at 35° C at the lower non-inhibitory salt amounts, and in one instance also at 25° C. NON-DORM was sensitized to greater germination in the dark than in the light by salt: at 15° C at the higher non-inhibitory salt amounts, and at 25° C at the highest reported inhibitory salt concentration.

## DISCUSSION

Seed germination of all phenotypes inhibited by 2% or more of NaCl. The effects of lesser amounts of NaCl on each of the three phenotypes was highly dependent on the specific temperature and light conditions. Regardless of non-isogenic germplasm used, the three test phenotypes encompassed complete range of seed germination in a wide range of temperatures (15-35° C) and light conditions (+, -) in water: dormant, 4%; after-ripened, 26%; nondomrant, 90%. The provided a good range to detect responses to salinity over a wide range of germination phenotype expressions. The set of dormant, after-ripened and non-dormant *S. faberi* populations and phenotypes observed in these experiments provided an ideal group of seeds with which to compare the effects of salt, temperature and light over a wide range of germinability responses (0-98.3%). This full range of germinability states and responses represented the full range observable in seeds in natural soil seed pools, allowing the observation of both stimulatory and inhibitory responses.

The process of after-ripening in water sensitized dormant population 3781 germination to both temperature and light conditions, a response not observed in dormant and nondormant phenotype germination in water. After-ripening in population 3781 without salt narrowed the range of temperatures within which germination occurred to 25-35° C, increased germination in the light compared to dark at 35° C, and increased germination at 35° C relative to that at 25° C. In the presence of salt, after-ripening also sensitized after-ripened phenotype germination to both temperature and light conditions in a similar manner as occurred in water, a response not observed in dormant seeds. At the highest salt concentrations these temperature and light after-ripening effects where not observed in after-ripened seed, possibly due to the confounding effects of low germination by salt inhibition. Long-term after-ripening in dry, cool conditions of did sensitize nondormant seed germination to light conditions and temperature conditions, but only in a few instances. Nondormant germination was greater in the dark than in the light at 15° C (0.5-1% NaCl) and 25° C (1.5% NaCl) when compared within common salt and temperature conditions, the opposite and more limited sensitization than that in those after-ripened.

In most instances germination of dormant was not affected by temperature within common salt and light conditions. There was a tendency for germination at 25° C in the dark to be greater than that at 15° C, and in one instance better than at 35° C. In after-ripened seeds the same pattern of increased germination with increased temperature occurred in both light conditions, but was somewhat dampened in the dark. In the light, 35° C greater than 25° C greater than 15° C, in non-inhibitory amounts of salt. In the dark, 35° C greater than 15° C. The same pattern of increased germination with increased temperature occurred in both light conditions, but was somewhat dampened in the dark.In nondormant seeds temperature either had no effect in common salt and light conditions, or germination at 25° C in the light was greater than at 15° C, and at 35° C greater than at 15° C. Many instances of common salt and light conditions no affect of temperature on germination. Temperature dependent germination apparent in some instances in the presence of light wherein it was greater at 25° C than at 15° C, and some fewer in which it was greater at 35° C than at 15° C.

Dormant population 3781 germination was insensitive to light, when compared within common salt and temperature conditions. In most instances of common salt and light conditions, DORM germination was insensitive to light, but in some others there was a tendency for greater germination at 25° C than at 15° C. After-ripening of population 3781 sensitized after-ripened seeds to light-stimulated germination while it completely de-sensitized it to germination at 15° C and sensitized it to germination at 35° C. The results indicate, while 35° C-light is the maximum, that it might be more heat than just light. Why? Because 35° C in light was first, but second was 35° C dark, followed by 25° C light. Why does after-ripening of dry dormant seed in the dark at 4° C lead to such a profound increase in germination at warmer (35° C) in the light, in water as well as all non-inhibitory levels of NaCl? Light doesn't provide more oxygen, or bypass the restrictions to oxygen uptake that regulate *S. faberi* seed dormancy-germinability. Therefore, how does light-stimulated germination cause so many more seeds to germinate? Dark, cool, moist incubation sensitizes highly dormant seed to light and high temperature stimulated germination. This study duplicates the field conditions dormant and after-ripened seeds would experience after autumn abscission in the frozen winter, cool moist spring, and warming soil prior to germination. These ecological conditions are those that *S. faberi* seed dormancy-germinability traits have been selected in: after dispersal in freezing soil; early spring thawing saturation with oxygen in moist cool dark conditions; later in spring the introduction of light and heat with spring tillage and soil disturbance (Dekker, 2014a). These are robust signals for ideal times to germinate and resume growth and reproduction. Insensitivity to cool, and increased sensitivity to high, temperatures that are induced by after-ripening also ensures seedling emergence will be delayed until soil temperatures rise, restricting very early season recruitment. The morphological locus and physiological mechanism(s) by which *S. faberi* seed are sensitized by light is unknown.

## ACKNOWLEDGEMENTS

Authors would like to acknowledge Prof. Mike Freeling, University of California, Berkeley, Department of Plant and Microbial Biology for creating an environment conducive to observing the complex interactions reported herein; for his interest in salt-tolerant *Setaria* spp. genotypes thriving along the salty shores of southern Japan; and for recognizing the potential of weedy *Setaria* genotypes for crop improvement of their close relative *Oryza* spp. by exploiting genomic synteny. We also thank Dr. Kari Jovaag for analyzing this complex data set.

**Table 1**. Effect of salt (percent NaCl, weight to volume (W:V)), temperature (° C) and light (light, dark) on seed germination (%) of dormant lot 3781 (no after-ripening; DORM), after-ripened (21 D) lot 3781(AFTER-RIPE), and non-dormant lot 1816 (NON-DORM) genotypes and phenotypes; means in the same column with the same letter are not significantly different with the Tukey mean separation test; means in the same row in bold are significantly different from row means in non-bold with the Tukey mean separation test.

|  |  |  | --------Genotype/Phenotype Germination (%)-------- | | |
|---|---|---|---|---|---|
| NaCl | Temperature |  | DORM | AFTER-RIPE | NON-DORM |
| (% W:V) | (° C) | Light/Dark | Lot 3781 | Lot 3781 | Lot 1816 |
| 0 | 15° C | light | 1.9 CDEF | 0 J | **79.4** ABCD |
| 0 | 15° C | dark | 3.9 ABCDEF | 0 J | **90.3** ABCD |
| 0 | 25° C | light | 3.9 ABCDEF | 31.7 DE | **97.2** A |
| 0 | 25° C | dark | 8.9 ABCD | 18.3 EFGHIJ | **98.3** A |
| 0 | 35° C | light | **2.2** CDEF | 73.3 AB | 85.8 ABCD |
| 0 | 35° C | dark | 2.5 BCDEF | 34.2 DE | **87.8** ABCD |
| 0.01 | 15° C | light | 0 F | 0 J | **80.6** ABCD |
| 0.01 | 15° C | dark | 7.2 ABCDEF | 0 J | **93.1** ABC |
| 0.01 | 25° C | light | 6.7 ABCDEF | 20.8 DEFGH | **97.8** A |
| 0.01 | 25° C | dark | 8.6 ABCDE | 23.3 DEFG | 95.8 AB |
| 0.01 | 35° C | light | **1.4** CDEF | 78.3 A | 89.7 ABCD |
| 0.01 | 35° C | dark | 0 F | 28.3 DE | **93.1** ABC |
| 0.05 | 15° C | light | 0 F | 0 J | **74.4** CD |
| 0.05 | 15° C | dark | 6.4 ABCDEF | 0 J | **91.4** ABCD |
| 0.05 | 25° C | light | 8.3 ABCDE | 38.3 CD | **97.8** A |
| 0.05 | 25° C | dark | 11.7 A | 18.3 EFGHIJ | **91.9** ABCD |
| 0.05 | 35° C | light | **1.1** DEF | 70.8 AB | 95.6 AB |
| 0.05 | 35° C | dark | 1.1 DEF | 34.2 DE | **87.2** ABCD |
| 0.10 | 15° C | light | 0.8 DEF | 0 J | **75.8** BCD |
| 0.10 | 15° C | dark | 2.2 CDEF | 0 J | **90.6** ABCD |
| 0.10 | 25° C | light | 6.1 ABCDEF | 30.0 DE | **92.8** ABC |
| 0.10 | 25° C | dark | 10.8 AB | 15.8 EFGHIJ | 95.3 AB |
| 0.10 | 35° C | light | **5.3** ABCDEF | 62.5 AB | 89.7 ABCD |
| 0.10 | 35° C | dark | 3.9 ABCDEF | 25.8 DEF | **89.7** ABCD |
| 0.50 | 15° C | light | 0 F | 0 J | **38.9** F |
| 0.50 | 15° C | dark | 0.6 EF | 0 J | **78.9** ABCD |
| 0.50 | 25° C | light | 6.7 ABCDEF | 27.5 DE | **91.1** ABCD |
| 0.50 | 25° C | dark | 9.4 ABC | 15.8 EFGHIJ | **93.9** ABC |
| 0.50 | 35° C | light | **0** F | 55.0 BC | 84.2 ABCD |

| | | | | | | | | |
|---|---|---|---|---|---|---|---|---|
| 0.50 | 35° C | dark | 0.8 | DEF | 30.8 | DE | **90.8** | ABCD |
| | | | | | | | | |
| 1.00 | 15° C | light | 0 | F | 0 | J | 9.2 | H |
| 1.00 | 15° C | dark | 0 | F | 0 | J | **51.4** | EF |
| 1.00 | 25° C | light | 1.4 | CDEF | 5.0 | GHIJ | **80.3** | ABCD |
| 1.00 | 25° C | dark | 8.1 | ABCDEF | 17.5 | EFGHIJ | **91.7** | ABCD |
| 1.00 | 35° C | light | 0 | F | 7.5 | FGHIJ | **84.2** | ABCD |
| 1.00 | 35° C | dark | 1.1 | DEF | 20.0 | DEFGHI | **92.5** | ABC |
| 1.50 | 15° C | light | 0 | F | 0 | J | 1.1 | H |
| 1.50 | 15° C | dark | 0 | F | 0 | J | 13.6 | GH |
| 1.50 | 25° C | light | 0.8 | DEF | 0 | J | **41.7** | F |
| 1.50 | 25° C | dark | 0.6 | EF | 4.2 | HIJ | **71.7** | DE |
| 1.50 | 35° C | light | 0 | F | 5.8 | GHIJ | 16.7 | GH |
| 1.50 | 35° C | dark | 0 | F | 0.9 | IJ | 33.9 | FG |